# STUDY OF SECOND AND THIRD HARMONIC GENERATION FROM AN INDIUM TIN OXIDE NANOLAYER: INFLUENCE OF NONLOCAL EFFECTS AND HOT ELECTRONS

Laura Rodríguez-Suné,[1,*] Michael Scalora,[2] Allan Johnson,[3] Crina Cojocaru,[1] Neset Akozbek,[4] Zachary Coppens,[4] Daniel Perez-Salinas, [3] Simon Wall,[3] and Jose Trull[1]

[1]Department of Physics, Universitat Politècnica de Catalunya, 08222 Terrassa, Spain
[2]Charles M. Bowden Research Center, CCDC AVMC, Redstone Arsenal, AL 35898-5000 USA
[3] ICFO – The Institute of Photonics Sciences, The Barcelona Institute of Science and Technology, 08860 Castelldefels (Barcelona), Spain
[4]AEgis Technologies Inc., 401 Jan Davis Dr., Huntsville, Alabama 35806, USA
*laura.rodriguez.sune@upc.edu

**Abstract:** We report comparative experimental and theoretical studies of second and third harmonic generation from a 20nm-thick indium tin oxide layer in proximity of the epsilon-near-zero condition. Using a tunable OPA laser we record both spectral and angular dependence of the generated harmonic signals close to this particular point. In addition to the enhancement of the second harmonic efficiency close to the epsilon-near-zero wavelength, at oblique incidence third harmonic generation displays unusual behavior, predicted but not observed before. We implement a comprehensive, first-principles hydrodynamic approach able to simulate our experimental conditions. The model is unique, flexible, and able to capture all major physical mechanisms that drive the electrodynamic behavior of conductive oxide layers: nonlocal effects, which blueshift the epsilon-near-zero resonance by tens of nanometers; plasma frequency redshift due to variations of the effective mass of hot carriers; charge density distribution inside the layer, which determines nonlinear surface and magnetic interactions; and the nonlinearity of the background medium triggered by bound electrons. We show that by taking these contributions into account our theoretical predictions are in very good qualitative and quantitative agreement with our experimental results. We show that by taking these contributions into account our theoretical predictions are in very good qualitative and quantitative agreement with our experimental results. We expect that our results can be extended to other geometries where ENZ nonlinearity plays an important role.

**Introduction**

Harmonic generation is one of the most fundamental processes in nonlinear optics that has been widely used for new coherent light source generation. Second and third harmonic generation (SHG and THG) have been studied extensively in different optical materials. High conversion efficiencies in traditional nonlinear optical devices usually require thick nonlinear materials with large nonlinearities, phase-matching conditions and low material absorption at the fundamental and harmonic wavelengths. If we focus on the particular case of harmonic generation at the nanoscale, which we strictly interpret to mean that material features may be only a few atomic diameters in size, then SH and TH efficiencies may decrease compared to macroscopic counterparts. On the other hand, phase matching conditions and even absorption may no longer play a primary or significant role. At the same time, the combination of geometrical features, resonances, bound electrons, and nonlocal, surface and magnetic effects on free electrons can substantially modify the material response and the very nature of linear and nonlinear light-matter interactions. At the nanometer scale conventional approximations to the dynamics of light-matter interactions break down. Therefore, new strategies must be sought in order to study, understand and ultimately harness the performance of sub-wavelength nonlinear optical materials, which nowadays are routinely produced and integrated in different devices and applications.

Another promising way to improve the performance of nonlinear optical devices is provided by a new class of optical materials that display vanishingly small real part of the dielectric constant, known as epsilon-near-zero (ENZ) materials [1-13]. These materials enhance the local electromagnetic field through the condition where the longitudinal component of the displacement vector of a TM-polarized field has to be continuous across a boundary between media with different optical properties. For homogeneous, flat structures this condition may be written as: $\varepsilon_{in} E_{in}^z = \varepsilon_{out} E_{out}^z$; $\varepsilon_{in}$ and $\varepsilon_{out}$ are the dielectric constants inside and outside the medium, respectively; $E_{in}^z$ and $E_{out}^z$ are the corresponding longitudinal components of the electric field amplitude, and require oblique incidence to excite the ENZ point. Therefore, if $\varepsilon_{in}$ decreases, then $E_{in}^z$ increases and nonlinear optical phenomena are enhanced, including nonlinear index of refraction [1], harmonic generation, optical bistability, and soliton excitation [2-13].

While ENZ materials can be made artificially, all natural bulk materials that display a Lorentz-like response also exhibit a real part of the dielectric permittivity that crosses zero, in proximity of either plasma or interband transition frequencies [2]. For instance, semiconductors like GaAs, GaP and Si display ENZ conditions near 100nm, deep in the ultraviolet range. Metals like Au, Ag and Cu have ENZ crossing points in the visible range, while the zero-crossing points of conducting oxides like indium tin oxide (ITO) and cadmium oxide (CdO) fall in the infrared regime and can be tuned using thermal post-processing of sputtered samples. The main limitations for field enhancement in ENZ materials are thought to be related to the narrow-band nature of the crossing "point", and absorption, i.e. the imaginary part of the dielectric constant, which naturally and significantly mitigates any potential singularity. For example, if we represent the boundary condition using complex variables, then when $\mathrm{Re}(\varepsilon_{in}) = 0$ the complex local field inside the material becomes $E_{in}^z = -iE_{out}^z / \mathrm{Im}(\varepsilon_{in})$. Causality thus removes the possibility of diverging denominators.

ITO is one of the most studied transparent conductive films for its practical applications in flat panel touch displays, aircraft windows for defrosting, and PV cells. It is a free-electron system characterized by absorption that is typically much smaller compared to that of noble metals, especially near the ENZ point. These dispersive properties have consequences that are important because they can trigger novel, low-intensity nonlinear optical phenomena that can usually be observed only for high local fields. Therefore, the study of ENZ materials can shed new light on our fundamental understanding of the optical properties of so-called free-electron systems at the nanoscale.

The linear and nonlinear optical properties of ITO and CdO [14, 15] are under intense scrutiny because of a readily available ENZ response tunable across the near infrared wavelength range. While metals and conducting oxides are both free electron systems, their linear and nonlinear optical properties do not completely mirror each other. The crucial physical differences between noble metals and conducting oxides are that: (1) free electron densities of the former can be several orders of magnitude larger than the free electron densities of the latter. This leads to significant differences in field penetration and the excitation of surface and volume nonlinearities, and nonlocal effects [16]; (2) interband excitations, increased free carrier density and a dynamic blueshift of plasma frequency characterize noble metals; conducting oxides are prone to display intraband transitions and increased electron gas temperature that lead to increased effective electron mass and a dynamic redshift of the plasma frequency [17]. Compared to noble metals, these effects make conducting oxides intriguing for the experimental and theoretical study of nonlinear optical interactions.

Recently, several studies of SHG and THG from ITO layers have been reported, most of them based on experimental observations. For instance, in reference [10] the ENZ crossing point was exploited for the enhancement of THG in a Kretschmann configuration. While detailed, the model used to simulate the experimental results utilized an effective but dispersionless $\chi^{(3)}$ having no specified origin. In reference [11], THG was reported for an ITO nanolayer, along with evidence that enhancement of the generated signal came as a result of an ENZ crossing point. Notwithstanding the absence of detailed numerical simulations, the nonlinearity was attributed entirely to the background crystal, with a claimed efficiency 600 times larger compared to that of crystalline bulk silicon. In reference [12], a comparative study of ITO and TiN nanolayers showed an enhancement of SHG from ITO at the ENZ wavelength. Effective surface nonlinearities were postulated without the benefit of a detailed analysis of the microscopic origins of SHG, or any other dynamical factors. Finally, the authors of references [1] and [9] concluded that in their experiments third and higher order nonlinearities were triggered by heating of the free electron cloud, once again without the benefit of a detailed, microscopic quantitative model. For an exhaustive list of experimental results we refer the reader to reference [9].

As we have seen, despite the reported experimental evidence to date no simultaneously comprehensive and satisfactory theoretical model has been presented, that sheds light on the competing physical mechanisms that characterize the interaction. For instance, it is known that nonlocal effects tend to blueshift the plasmonic resonance [14-16]. In contrast, absorption, which is in turn modified by nonlocal effects, can change the free electron effective mass, causing the plasma frequency to redshift [17]. Nonlocal effects and modulation of the plasma frequency are dynamic, time dependent factors that can strongly influence the propagation, and can easily combine to impress unique dynamical features on measurable quantities, such as spatial and temporal modulation of dielectric constant and refractive index. Nevertheless, most models rely either on the introduction of phenomenological or effective parameters, or on qualitative, seemingly plausible arguments that may help one glimpse the nature of the interaction. However, in the absence of detailed theoretical models to fully describe propagation and light-matter interaction phenomena, these arguments have the potential of not only being inaccurate but also lead to misinterpretations, thus limiting one's understanding of the underlying physical mechanisms. In a centrosymmetric medium SHG depends strictly on the effective electron mass, charge density distribution, magnetic coupling, nonlocal effects, and to some extent frequency downconversion. As we will see, the source of THG and/or nonlinear index modulation remains somewhat ambiguous, since both the background lattice and the free electron cloud can contribute to it.

In view of our discussion above, in what follows we first present experimental measurements of SHG and THG from a 20nm-thick ITO layer as a function of the incident laser pulse wavelength in order to demonstrate and quantify the resonance-like properties that occur at the ENZ spectral wavelength. We then go on to show the angular dependence of SHG for different input wavelengths in order to highlight the metal-like response, an aspect which is often neglected in conducting oxides and that typically yields maximum SHG at large angles. In order to proceed unencumbered we will describe the most salient but necessary aspects of the theoretical model, and compare with experimental results. We leave the intricate details of the model for a later publication as it is beyond the scope of this work.

**Experimental set-up and results**

The ITO nanolayers used in our experiments are RF magnetron sputtered from an ITO target ($In_2O_3/SnO_2$ 90/10wt %, purity 99.99%) on transparent fused silica glass. The electro-optical properties of deposited ITO films depend on a number of processing conditions, including substrate temperature, RF sputtering power, deposition pressure, oxygen partial pressure, and post-annealing. The samples were fabricated at room temperature with a RF power of 100W and Ar sputter pressure of 5mTorr. The *as-is* sputtered ITO typically has an ENZ crossing above 1800nm. The samples were post-annealed in vacuum so that a zero-crossing of the real part of the dielectric constant is blueshifted as electron density and/or electron mobility increase. The samples were analyzed using a variable angle spectroscopic ellipsometer (Woollam, VASE). Using a Lorentz-Drude dispersion model, complex dielectric constant and film thickness were retrieved. Additional, angular p-polarized transmission measurements were made to verify the spectral location of the ENZ crossing point. The resistivity of ITO is attributed to oxygen vacancy and film crystallinity. Therefore the deposition conditions as well as post-annealing process allow one to tune the ENZ wavelength [11, 12].

Our 20-nm thick sample was annealed at 600ºC and displays an ENZ condition near 1260nm. It has been shown that the electro-optical properties of ITO can be thickness-dependent even under the same deposition conditions, and thus may exhibit a different ENZ condition. This fact immediately introduces some uncertainty in the retrieved dielectric constant, because one possible interpretation of this apparent discrepancy for similarly grown and treated samples is that the data of very thin layers may intrinsically contain shifts due to nonlocal effects. Nonlocal effects are strongest near surfaces and are determined by the magnitude of the Fermi energy, $E_F = \frac{\hbar^2}{2m^*}\left(3\pi^2 n_0\right)^{2/3}$, which in turn depends on the effective electron mass and the carrier density. In our simulations we leverage the well-known fact that effective mass and density can significantly change during post-growth thermal processing [18]. Perhaps more importantly, it also suggests that if nonlocal effects are important, data that is retrieved for a given layer thickness should probably not be used to model a layer having different thickness. We will expand on this and other issues in a separate work.

The spectral measurements of second and third harmonic signals generated by the ITO sample were performed using a commercial OPA (TOPAS, Light Conversion) pumped by 4 mJ pulses from a Ti:sapphire amplifier (Astrella, Coherent) delivering continuously-tunable pulses in a wavelength range from 1140 nm to 1600 nm, with energies exceeding 0.5mJ/pulse generated at a repetition rate of 1kHz. A long-pass filter (Thorlabs FELH1100) was used to remove the residual pump and other sum-frequency components coming from the OPA wire grid polarizer and a super-achromatic waveplate was employed to prepare and control the input beam polarization. The pulses were attenuated down to tens of μJ using reflective, neutral density filters. For the characterization of the input beam at the plane of the sample, knife-edge and FROG measurements were carried out. Depending on the wavelength, beam waists between 4.2 mm and 4.8 mm at $1/e^2$ radius in intensity and pulse durations between 50 and 102fs FWHM, leading to typical power densities between 1 and 2 GW/cm$^2$ were measured. For the SHG measurements, the fundamental beam was directed onto the sample under these conditions while in the case of the THG measurements a 50 cm, uncoated CaF$_2$ lens was introduced in the beam path, leading to a focal spot size of 145μm at 1240nm. In order to avoid possible harmonic signals arising from portions of the setup placed before the ITO sample were filtered out of the beam path using band-pass filters. The ITO sample was mounted on a rotating stage ensuring a precise control of the fundamental beam incident angle. A schematic representation of the set-up is shown in Fig.1. After the sample, the light was collimated by a lens and the polarization of the generated SH or TH was analyzed by a polarizer. We expect to detect faint SH and TH signals (typical efficiencies of the order $10^{-8}$-$10^{-9}$). It is then crucial to separate them completely from the fundamental beam and to avoid any possible harmonic generation arising from other surfaces of optical elements placed after the ITO sample. To this end, different filters were used to attenuate the fundamental beam immediately after the sample. A prism and a sharp edge were used to separate and obscure the remaining fundamental radiation from the SH or TH path. A photomultiplier tube (Hamamatsu) was used to detect the harmonic signals together with a spectral filter having 20nm band pass transmission band around either the SH or TH frequency. The entire detection system was completely darkened and mounted on a rotating rigid platform, allowing measurement of either transmittance (as represented in Fig.1) or reflectance.

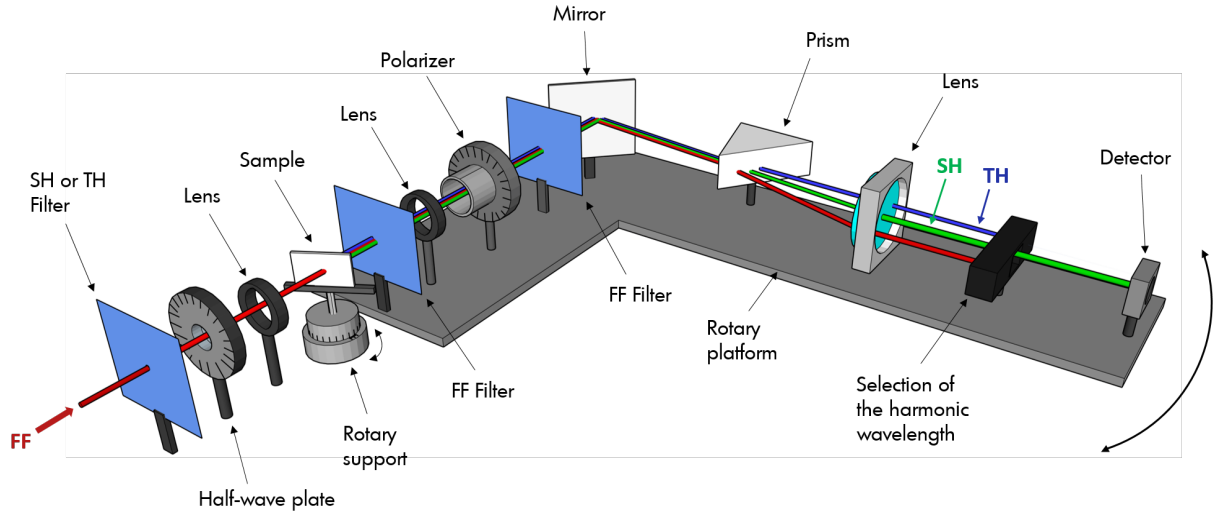

**Figure 1:** Setup for measuring SH and TH signals. The whole platform after the sample is mounted on a rotating arm to be used for transmission or reflection measurements.

A detailed calibration procedure was performed in order to estimate accurately the efficiencies of the process as the ratio between the generated SH or TH intensity in the scattered pulses (transmitted or reflected) and the total initial peak pump pulse intensity. To this end, we measured separately the energy and pulse durations at each particular wavelength used in the experiments, and we substituted the ITO sample by a BBO crystal, generating a larger SH signal that could be measured with a calibrated photodiode and later with the photomultiplier, after being attenuated using neutral density filters. This procedure allows us to precisely relate the signal measured by the photomultiplier to that generated at the sample plane. This experimental setup allows the recording of harmonic conversion efficiencies as low as $10^{-10}$.

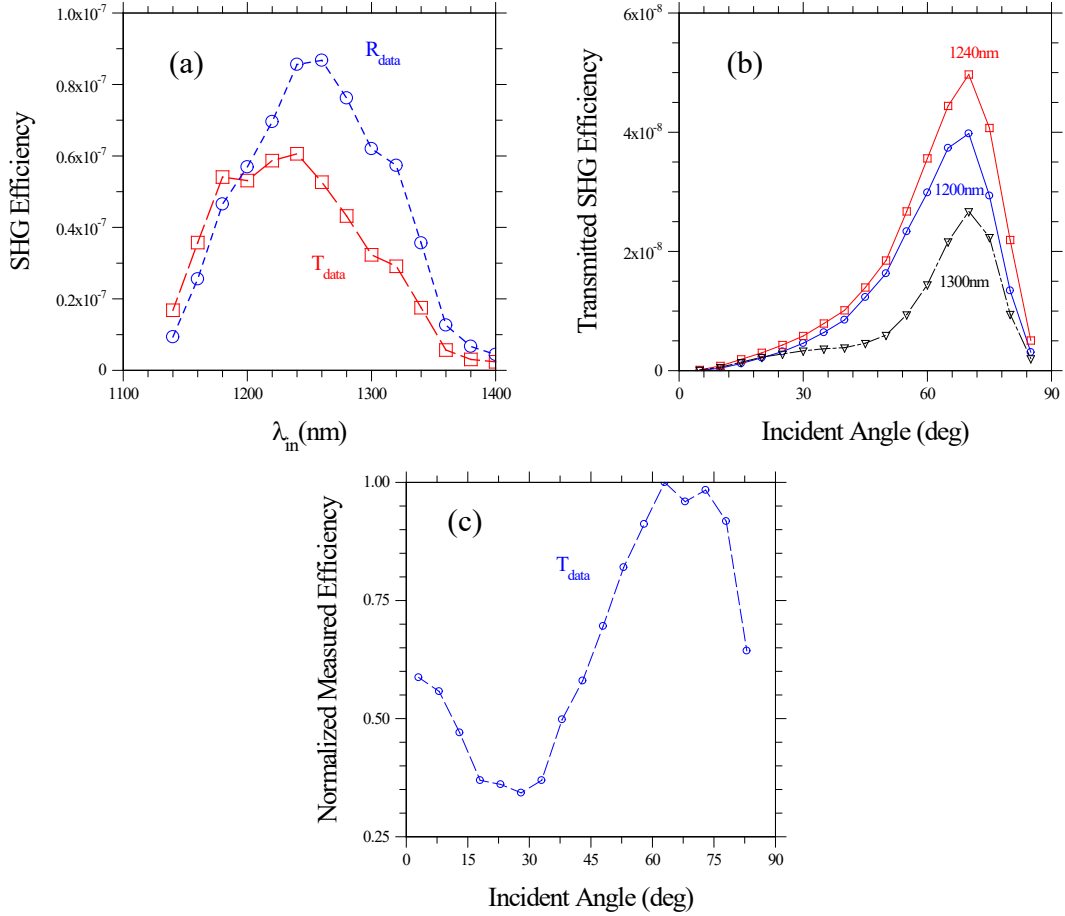

**Figure 2:** (a) Experimental data of reflected (dashed blue curve, open circle markers) and transmitted (dashed red curve, open square markers) SHG efficiency from a 20nm-thick ITO layer vs. input carrier wavelength. (b) Angular dependence of transmitted SHG vs incident angle at the indicated carrier wavelengths. The interesting feature in this figure is the shoulder that develops below 40° in the conversion efficiency at 1300nm. (c) Normalized transmitted THG conversion efficiency vs. incident angle for incident carrier wavelength of 1240nm.

In the first experiment we recorded the SH efficiency as a function of the fundamental wavelength around the ENZ condition, for a fixed incident angle of 60°. The spectral response of the SH generated both in transmission (dashed red curve, open squares) and in reflection (dashed blue curve, open circles) is shown in Fig.2 (a). The curves show a clear spectral dependence of the SHG conversion efficiency and its enhancement near the ENZ condition, appearing near 1240nm. In this regard, it is important to note that the locations of reflected and transmitted maxima clearly do not coincide, most likely because transmission hinges more on propagation through the bulk material, while the reflected component partially benefits from conditions at the surface(s). This relative shift between transmitted and reflected maxima was previously unknown. In a second experiment we fixed the fundamental wavelength and we recorded the angular dependence of the SH efficiency. Fig.2 (b) shows three angular dependence curves of SHG at three different fundamental wavelengths: above (1300nm), at (1240nm) and below (1200nm) the ENZ condition. Finally, in Fig.2 (c) we show the normalized THG conversion efficiency, measured in transmission, as a function of incident angle at the carrier wavelength of 1240nm. We normalize THG because, as we will see below, the contribution of the bound and free electron contribution to the source of TH light at the moment remains somewhat ambiguous. The most salient feature of the TH signal is the minimum that occurs near 30º, an effect that had previously been theoretically predicted for an ITO/Au bilayer when light was incident from the ITO side [19].

**Theoretical approach**

As already mentioned above, an approach where the mere effective linear and/or nonlinear optical properties are pursued, or perhaps where just the magnitude of the nonlinear index change is sought using a unidirectional beam propagation method with an effective nonlinearity of unspecified origin, can limit our understanding of intricate and subtle processes, primarily because an effective response ignores surfaces and obscures internal details of the interaction. For this reason we adopt a theoretical approach that embraces full scale time-domain coupling of matter to the macroscopic Maxwell's equations. Our strategy is based on formulating a microscopic, hydrodynamic approach where the aim is to understand linear and nonlinear optical properties of conducting oxides by accounting for competing surface, magnetic, and bulk nonlinearities arising from both free and bound charges, preserving linear and nonlinear dispersions, nonlocal effects due to pressure and viscosity, changes in the effective mass of free electrons triggered by absorption, and even pump depletion, should local field intensities be so large to require it.

The theoretical description of macroscopic electrodynamics of conducting oxides found in the literature typically begins and ends with a simple Drude model given in Eq.(1):

$$\varepsilon_{ITO}(\omega) = \varepsilon_\infty - \frac{\omega_p^2}{\omega^2 + i\gamma_f \omega}. \qquad (1)$$

The magnitude of $\varepsilon_\infty$ may depend on thermal post-processing of the sample, it is usually given as constant, and is not unique because it depends on the crossing point. In Eq.(1), $\omega_p^2 = \frac{4\pi n_{0,f} e^2}{m_f^*}$ is the plasma frequency; $n_{0,f}$ the free electron density; $e$ the electronic charge; $\gamma_f$ the free electron damping coefficient; and $m_f^*$ the temperature-dependent effective free electron mass. Strictly speaking, referring to an "$\varepsilon_\infty$" other than unity is improper because as $\omega \to \infty$ the intervening volume does not contain any matter. In reality, this contribution is triggered by a background and quite observable bound electron resonance having a Lorentzian response peaked somewhere in the UV range, such that when combined with the Drude contribution as follows:

$$\varepsilon_{ITO}(\omega) = 1 - \frac{\omega_{p,b}^2}{\omega^2 - \omega_{0,b}^2 + i\gamma_b\omega} - \frac{\omega_p^2}{\omega^2 + i\gamma_f\omega}, \qquad (2)$$

the real part of the *total* dielectric constant of ITO crosses zero, with "$\varepsilon_\infty$"
$= 1 - \frac{\omega_{p,b}^2}{\omega^2 - \omega_{0,b}^2 + i\gamma_b\omega}$. Here, $\omega_{p,b}$ is the bound electron plasma frequency, defined similarly to the free electron counterpart; $\omega_{0,b}$ is the resonance frequency; and $\gamma_b$ the bound electron damping coefficient.

The portrayal of the local dielectric constant $\varepsilon_{ITO}(\omega)$ as in Eq.(2) immediately changes the picture and adds a pivotal, dynamical dimension that is usually completely disregarded: the Lorentzian response triggers an intrinsic, dispersive, resonant nonlinearity that may compete with the nonlinearity introduced by heating the free carriers, depending on tuning and incident pump intensity. Another aspect that significantly limits the description of ENZ materials is that both Eqs.(1) and (2) also neglect to include the impact of nonlocal effects.

With these caveats we can write the coupled material equations of motion that describe two separate polarization components produced by free and bound electrons, without explicitly introducing the dielectric constant, or making any other assumption that would automatically assume singular denominators to explain larger-than-usual conversion efficiencies. Fitting the retrieved, local dielectric constant merely serves the purpose of determining damping coefficients, effective masses, and densities to be inserted in dynamical equations of motion from which a dielectric constant can be retrieved, should one wish to pursue it. Both equations simultaneously contain lowest order contributions of bulk, surface and magnetic nonlinearities, which we now adapt from our previous work on nonlinear optics of metals [14, 19-22] and semiconductors [23, 24], by modifying them to include the dynamics that ensues with a temperature and thus time-dependent plasma frequency:

$$\begin{aligned}
\ddot{\mathbf{P}}_f + \gamma_f\dot{\mathbf{P}}_f &= \frac{n_{0,f}e^2}{m_f^*(T_e)}\mathbf{E} - \frac{e}{m_f^*(T_e)}\mathbf{E}(\nabla\bullet\mathbf{P}_f) + \frac{e}{m_f^*(T_e)}\dot{\mathbf{P}}_f\times\mathbf{H} \\
&+ \frac{3E_F}{5m_f^*(T_e)}\left(\nabla(\nabla\bullet\mathbf{P}_f) + \nabla^2\mathbf{P}_f\right) - \frac{1}{n_{0,f}e}\left[(\nabla\bullet\dot{\mathbf{P}}_f)\dot{\mathbf{P}}_f + (\dot{\mathbf{P}}_f\bullet\nabla)\dot{\mathbf{P}}_f\right]
\end{aligned}, \qquad (3)$$

$$\ddot{\mathbf{P}}_b + \gamma_b \dot{\mathbf{P}}_b + \omega_{0,b}^2 \mathbf{P}_b + \mathbf{P}_{b,NL} = \frac{n_{0,b} e^2}{m_b^*} \mathbf{E} + \frac{e}{m_b^*} (\mathbf{P}_b \bullet \nabla) \mathbf{E} + \frac{e}{m_b^* c} \dot{\mathbf{P}}_b \times \mathbf{H} \; . \qquad (4)$$

Eqs.(3) and (4) describe the dynamics of free and bound electrons, respectively; $T_e$ is the free electron temperature; $\mathbf{P}_{b,NL} = \alpha \mathbf{P}_b \mathbf{P}_b - \beta (\mathbf{P}_b \bullet \mathbf{P}_b) \mathbf{P}_b + \dots$ is the bulk crystal's nonlinear polarization; $m_b^*$ is the bound electron mass; $n_{0,b}$ is the bound electron density. The coefficients $\alpha$ and $\beta$ are to be interpreted as tensors, and carry information about crystal symmetry. For example, $\alpha = 0$ for centrosymmetric media like ITO or noble metals. The third order nonlinearity $\beta (\mathbf{P}_b \bullet \mathbf{P}_b) \mathbf{P}_b$ reflects the symmetry of an isotropic medium, but can easily be generalized to describe other crystal symmetries. The combination of Eqs.(3) and (4) thus preserves both linear and nonlinear dispersions as well as surface harmonic generation. The free electron component in ITO is similar to the free electron polarization component typical of noble metals, except for the fact that $m_f^*$ is now a function of temperature. We emphasize that neither the dielectric constant nor the index of refraction are found in the dynamical equations of motion, and no assumptions are made regarding either quantity. Ultimately, Eq.(3) can also be adapted to describe a time-varying free electron density triggered by interband transitions in metal and/or semiconductors. In summary, the first and second term on the right hand side of Eq.(3) are Coulomb terms, followed by the magnetic Lorentz term, nonlocal terms (pressure and viscosity,) and ending with convective terms [14, 20]. On the right side of Eq.(4) we have the bound electron plasma frequency, followed by surface and magnetic Lorentz contributions [23, 24].

The description of hot carriers is typically done by implementing the two-temperature model [25, 26], which couples the lattice temperature to that of the electron gas and is used to determine the instantaneous plasma frequency. However, as in the case of silicon, if the electron temperature is only a few thousand degrees Kelvin one may assume that for ITO the following approximation is also valid [27]:

$$m_f^*(T_e) \approx m_0^* + \alpha K_B T_e = m_0^* + \alpha K_B \Lambda \iint \mathbf{J} \bullet \mathbf{E} \, d\mathbf{r}^3 dt \qquad , \qquad (5)$$

where $K_B$ is Boltzmann's constant; $\alpha$ and $\Lambda$ are constants of proportionality; $m_0^*$ is the electron's rest mass for no applied field; $\Lambda \iint \mathbf{J} \bullet \mathbf{E} \, d\mathbf{r}^3 dt$ represents absorption; and $\mathbf{J} = \dot{\mathbf{P}}_f$ is the current density derived from Eq.(3), which contains intrinsic modifications imparted to the dielectric response by nonlocal effects and hot electron contributions. However, for our present purposes and for convenience in what follows we will assume the simplified form $\mathbf{J} = \sigma_0 \mathbf{E}$ ,

where $\sigma_0$ is a constant to be determined. Using the temperature dependent expression for the effective mass Eq.(5), the leading term on the right hand side of Eq.(3) may be written as:

$$\frac{n_{0,f}e^2}{m_f^*(T_e)}\mathbf{E} = \frac{n_{0,f}e^2}{m_0^*}\left(1 + \frac{\Lambda}{m_0^*}\iint \mathbf{J} \bullet \mathbf{E}\, d\mathbf{r}^3 dt\right)^{-1}\mathbf{E} \approx$$
$$\frac{n_{0,f}e^2}{m_0^*}\left(1 - \frac{\Lambda\sigma_0}{m_0^*}\iint \mathbf{E} \bullet \mathbf{E}\, d\mathbf{r}^3 dt + \left(\frac{\Lambda\sigma_0}{m_0^*}\right)^2\left(\iint \mathbf{E} \bullet \mathbf{E}\, d\mathbf{r}^3 dt\right)^2 + ...\right)\mathbf{E} \qquad . (6)$$

The effective mass appears in all but one term on the right hand side of Eq.(3). However, additional nonlinear contributions beyond Eq.(6) to the rest of the terms in Eq.(3) are of higher order, and for our present purposes may be neglected. For simplicity we may also drop the volume integral in Eq.(6) and introduce a parameter $\tilde{\Lambda}$ proportional to the product of $\frac{\Lambda\sigma_0}{m_0^*}$, the interaction volume and the temporal duration of the pulse, so that Eq.(6) takes the following, simplified form:

$$\frac{n_{0,f}e^2}{m_f^*(T_e)}\mathbf{E} \approx \frac{n_{0,f}e^2}{m_0^*}\mathbf{E} - \tilde{\Lambda}\left(\mathbf{E} \bullet \mathbf{E}\right)\mathbf{E} + \tilde{\Lambda}^2\left(\mathbf{E} \bullet \mathbf{E}\right)^2\mathbf{E} \qquad (7)$$

The coefficient $\tilde{\Lambda}$ thus determines the spatio-temporal dynamics of the redshift impressed upon the plasma frequency. While this discussion leaves Eq.(4) unaltered, the first term on the right hand side of Eq.(3) takes the form of Eq.(7), and includes cubic and quintic nonlinearities. Finally, using the solutions of Eqs.(3), (4) and (7) the total polarization is written as the vector sum of free and bound electron contributions, $\mathbf{P}_{Total} = \mathbf{P}_f + \mathbf{P}_b$, and is inserted into Maxwell's equations.

**Experimental versus theoretical results**

For ease of reference, in Fig.3(a) we reproduce the transmitted (dashed red curve, open squares) and reflected (dashed blue curve, open circles) SHG spectra recorded as a function of input carrier wavelength, at the fixed incident angle of 60°, as in Fig.2(a). The solid curves labeled $T_{sim}$ (red) and $R_{sim}$ (blue) are the corresponding spectra calculated by integrating the material equations (3), (4) and (7) together with Maxwell's equations, using pulses that are plane in the transverse direction to remove effects due to diffraction, and are approximately 100fs in duration. If pulses are not strongly affected by spatial or temporal dispersion and do not change shape, as in our case, then conversion efficiencies may be defined as the ratio of either peak intensities or pulse energies interchangeably, with identical results. Fig.3 (b) contains a plot of the experimentally-retrieved, local complex dielectric constant. In this instance the real part of the dielectric constant crosses zero near 1260nm. The data is fitted using the Drude-Lorentz oscillator in Eq.(2). While the data may vary slightly from sample to sample, with somewhat different crossing points, the curve is nevertheless strongly indicative of the fact that material dispersion is Drude-like above 450nm, and Lorentz-like at smaller wavelengths. This should make it clear that an accurate rendition of third order effects can come about only if the dynamics of both free and bound electrons are captured simultaneously.

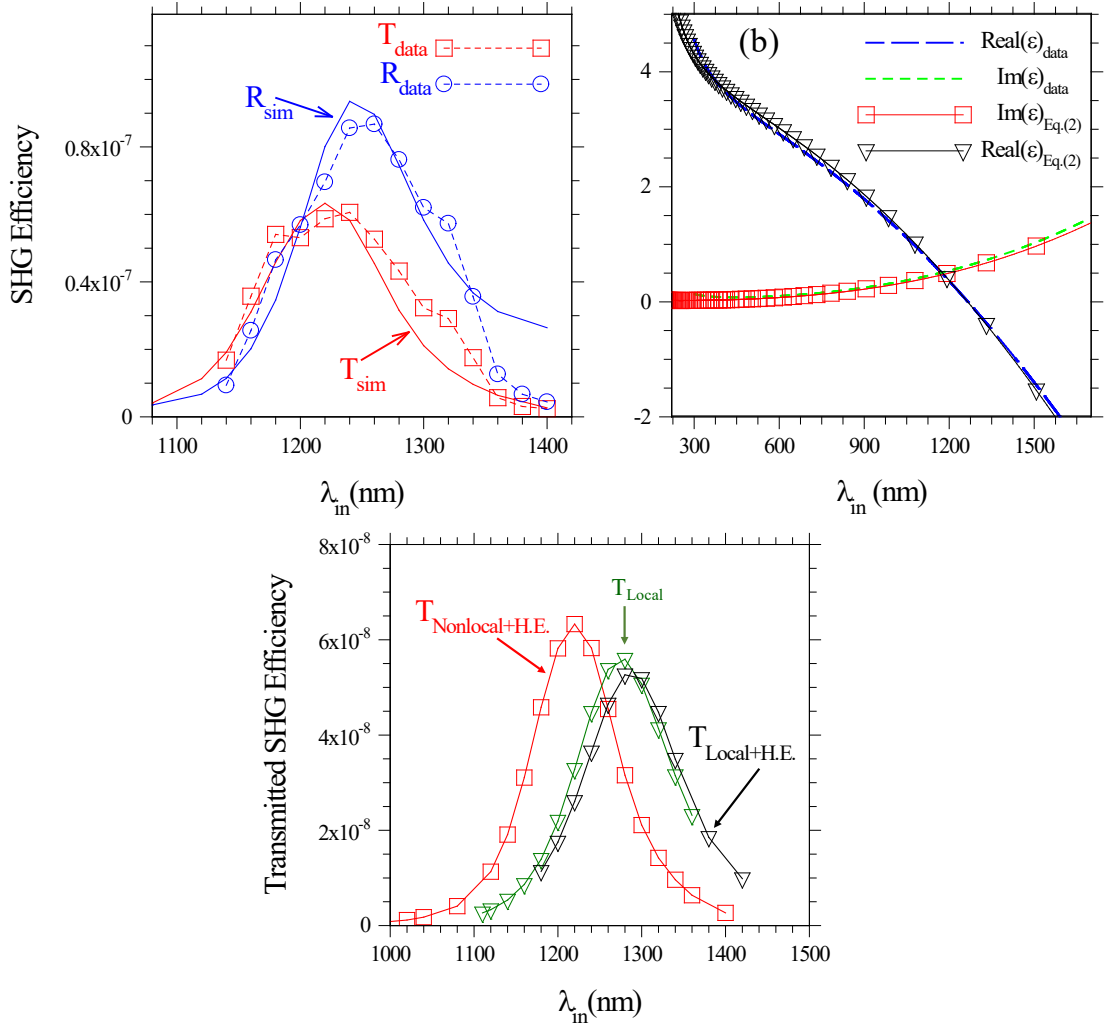

**Figure 3: (a)** Experimental measurement of reflected (dashed blue curve, open circle markers) and transmitted (dashed red curve, open square markers) SHG efficiency from 20nm-thick ITO layer vs. input carrier wavelength, as in Fig.2(a). The solid curves labeled $T_{sim}$ (red) and $R_{sim}$ (blue) are the corresponding theoretical predictions, as indicated by the respective arrows. **(b)** Data corresponding to real and imaginary parts of the dielectric constant (long blue dashes, and short green dashes, respectively) retrieved from ellipsometric analysis, fitted using the Drude-Lorentz oscillators in Eq.(2), as indicated in the legend. The real part of the dielectric constant crosses zero near 1260nm. **(c)** Transmitted SHG efficiency vs. input carrier wavelength. The subscript "H.E." refers to hot electrons. The green curve (solid tringle markers) labeled $T_{Local}$ is obtained by setting $E_F = 0$ and $\tilde{\Lambda} = 0$, i.e. without nonlocal effects or hot electrons. The black curve (open triangles) is redshifted by approximately 15nm with respect to the local curve, and is calculated with $E_F = 0$ and nonzero $\tilde{\Lambda}$. The red curve (open squares) labeled $T_{Nonlocal+H.E.}$ is blueshifted by approximately 35nm with respect to the local curve, and includes nonlocal and hot electron effects.

In Fig.3 (c) we plot only transmitted SHG conversion efficiencies to illustrate the contrasting actions of nonlocal and hot electron contributions. The figure suggests that at higher intensities plasma frequency redshift could entirely offset any blueshift triggered by nonlocal effects.

According to the structure of Eq.(3), increasing incident power density redshifts the plasma frequency and thus the ENZ condition, affecting SHG and THG simultaneously. The effective electron mass determines the nonlinear gain arising from Coulomb and magnetic Lorentz terms, while the density affects the nonlinear gain due to convection. By the same token, $m_0^*$ and $n_{0,f}$ determine the Fermi energy and thus the degree of blueshift the ENZ peak experiences [14]. The calculations are performed using a power density of $1\,\mathrm{GW/cm}^2$, but can be varied somewhat consistent with experimental conditions. The effective free electron mass and density are chosen as $m_0^* = 0.033 m_e$ and $n_{0,f} \sim 10^{20}/\mathrm{cm}^3$, which combine to yield a Fermi energy that imparts a blueshift of nearly 50nm to the SHG peak spectral position. The magnitude of $\tilde{\Lambda}$ is chosen so that its effect is to redshift the SHG spectra by approximately 15nm, for a net blueshift of ~35nm. The integration of the equations of motion is carried out using a spectral method to propagate the fields, and a predictor-corrector algorithm to integrate the material equations. The technique is described in our references [19-24].

From our discussion one may easily surmise that the reproduction of experimental conditions and results depends on a subtle balance between all these aforementioned factors, which are pivotal, are also at play in noble metals in similar fashion [19, 21], but whose magnitudes cannot be known precisely. Nevertheless, the agreement that we find between predictions and experiment is quite extraordinary, and extends to the relative locations of transmitted and reflected SH peaks, and the ratio between transmitted and reflected SH maxima. This ratio *is not* maintained if for instance we arbitrarily set to zero the magnetic Lorentz contribution (not shown), although the peaks remain similarly shifted with respect to each other. By the same token, artificially eliminating Coulomb and convective (surface) terms instead leads to a completely different SH response, an indication that one should be wary of using models that merely insert effective, dispersionless surface and/or volume nonlinearities in order to reproduce the shape of experimental curves, if not the amplitudes. As may be ascertained from Fig.3 (c), nonlocal effects and increased electron gas temperature tend to pull the resonance(s) in opposite directions. Since at larger intensities it becomes possible for the two effects to appear to offset each other, one might be tempted to simply ignore them. However, in reality both are present simultaneously and change the nature of the interaction in nontrivial ways. We will expand on this and other issues in a separate work that will focus on the model.

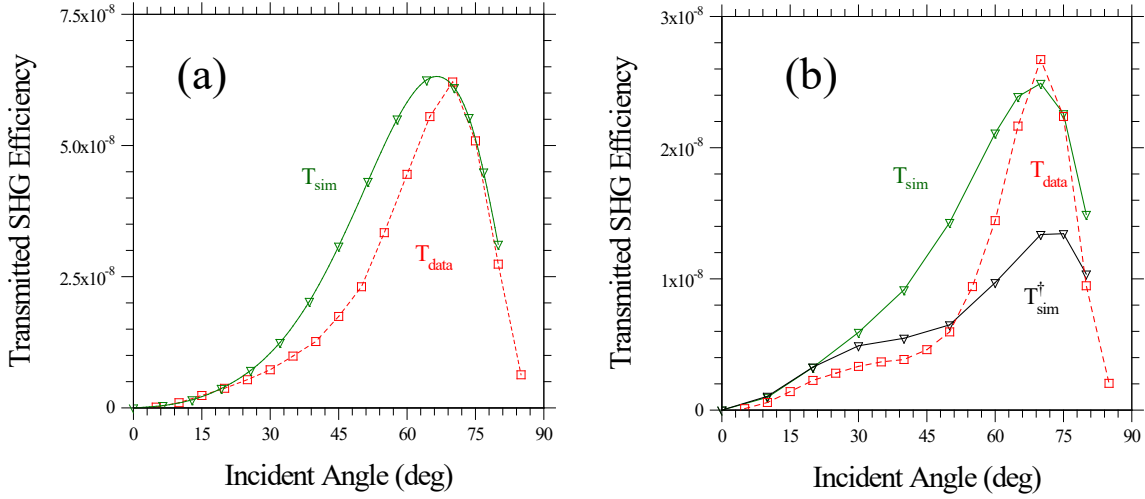

**Figure 4:** Measured (dashed red curve, empty squares, $T_{data}$, from Fig.2(b)) and predicted (solid green curve, full triangles, $T_{sim}$) transmitted SHG efficiency from 20nm-thick annealed ITO as a function of the angle of incidence for: (a) 1240nm, and (b) 1300nm. A shoulder can be seen to develop in the measured data at both wavelengths, but is much more pronounced at 1300nm. One way to reproduce the shoulder is by forcing the Fermi energy to be artificially large (solid black curve with empty triangles, $T_{sim}^{\dagger}$) in order to emphasize nonlocal effects without significantly impacting peak conversion efficiencies.

Features that may be ascribed to nonlocal effects are revealed by examining the angular dependence of SHG. In figure 4 we plot measured transmitted SHG conversion efficiencies at two fixed wavelengths as a function of incident angle, at and near the ENZ condition. Once again we find very good qualitative and quantitative agreement between our experimental observations and the predicted curves. The peak SH efficiencies, which occur near 70⁰, and the shape of the curves depend on the magnitude of the nonlocal coefficient, and to the extent it is counteracted by pump absorption and electron cloud heating. In Fig. 4(b) we plot SHG efficiency as a function of angle at 1300nm, which displays a clear shoulder below 40⁰. The predicted transmittance, represented by the blue curve labeled $T_{sim}$, displays a slight shoulder below 40⁰ before peaking at 70⁰. However, the black curve labeled $T_{sim}^{\dagger}$ can be obtained by either artificially increasing the Fermi energy, and thus the nonlocal coefficient, or by using a material dispersion curve that displays a slightly blueshifted ENZ crossing point, in either case without modifying either $m_0^*$ or $n_{0,f}$ so as not to affect SH gain. Given the evident and quite remarkable qualitative and quantitative agreement between our experimental observations and our predictions, in terms of both spectral response, peak locations, the ratio reflected/transmitted maxima, and angular dependence, including the ability to reproduce the shoulder shown in Fig.4(b), the model appears to clearly and accurately capture the most prominent aspects of the electromagnetic response of conducting oxides like ITO.

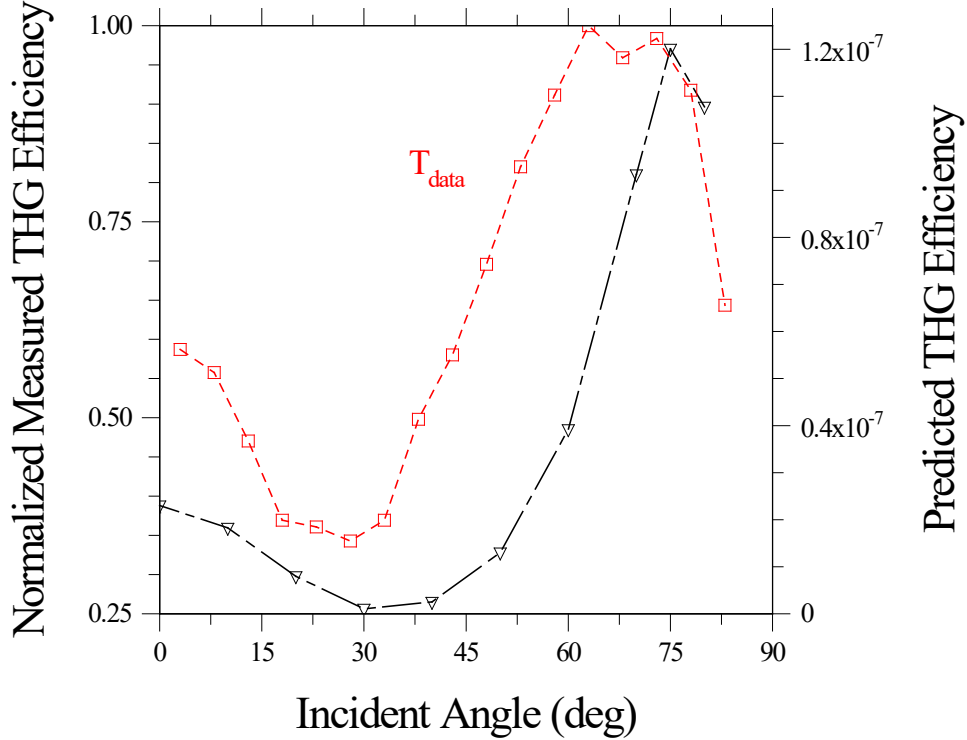

**Figure 5:** Transmitted THG from 20nm-thick annealed ITO as a function of the angle of incidence when the FF is tuned at 1240nm. Right axis: predicted THG efficiency. Left axis: Normalized measured THG efficiency, from Fig.2(c).

In Figure 5 we show the transmitted third harmonic efficiency as a function of incident angle when the carrier wavelength of the incident pump pulse is tuned to 1240nm. The predicted spectral response of THG (not shown) is generally similar but not identical to the spectral response of the SH signals, with transmitted and reflected peaks near the ENZ condition that are also shifted with respect to each other. We leave those details for a later work. The minimum that occurs at relatively small angles is a unique feature of the angular dependence of THG from an ITO layer that had been previously predicted to occur [19], but that presently remains unexplained. The measured TH conversion efficiency is plotted on the left axis and is normalized because its origin remains uncertain. In reference [11] THG was attributed entirely to the crystal's $\chi^{(3)}$. In references [1] and [9], the third order nonlinear response was assumed to arise entirely from the free electron cloud. In contrast, our equations (3) and (4) point to two dominant possibilities (although the free electron gas is also third, much weaker source of TH light via the Coulomb, magnetic and convection terms that we also take into account [20],) and make it possible to discern between them: the background crystal's $\chi^{(3)}$ (the coefficient $\beta$ in Eq.(4)) and hot electrons (the coefficient $\tilde{\Lambda}$ in Eq.(3)). These two types of nonlinearities behave differently because they are active in different wavelength ranges, tend to respond in different ways at large intensities, but give similar qualitative contributions at low intensities. Notwithstanding the possibility of THG having mixed origin, it is clear that the model once again captures remarkably well the dynamical aspects of the interactions. This feature of the model will also be discussed in details in a separate effort. Suffice it to say here that we have demonstrated its flexibility and its ability to accurately describe the qualitative and quantitative aspects of SHG and THG from a planar, conductive oxide structure.

**Conclusions**

In summary, we have presented experimental results on the spectral and angular response of second and third harmonic generation from a 20nm-thick ITO layer. The results suggest that transmitted and reflected SHG spectra display maxima that are shifted by approximately 20nm with respect to each other. This relative shift has not been reported before, and persists even in the absence of nonlocal effects and constant plasma frequency. Additional simulations suggest that this shift may depend on sample thickness. We have also explored the angular dependence of THG, and observed a minimum at approximately 30⁰ angle of incidence. Similar behavior has been predicted in the context of an ITO/Au bilayer, but has not been observed previously. We have also outlined a new theoretical approach based on a microscopic, hydrodynamic model that takes into account nonlocal effects, a time-dependent plasma frequency, surface and magnetic effects, convection and bound electron contributions that collectively give rise to surface and bulk generated harmonic signals in conducting oxides. In our case, maximum longitudinal field intensity amplification is a mere factor of two inside the layer across the ENZ resonance. This should be contrasted with what can be achieved in the context of nested plasmonic resonances, where local field intensities inside an ITO nanowire embedded inside a resonant, gold nano-antenna can be amplified in excess of three orders of magnitude [16]. Ultimately, our goal was to implement an accurate propagation model that took these physical factors into account in a realistic environment, without the manifestation of artificial, misleading singularities. To the best of our knowledge we are not aware of any other theoretical approach that is as comprehensive as outlined in this work, and that yields qualitatively and quantitatively accurate results in a context where the only free parameters are effectively free electron mass and density. In this regard, we expect our theoretical model to be applicable to other nonlinear interactions in ENZ materials and provide a deeper understanding of the phenomenon.

**Acknowledgment**

MS, NA, and ZC acknowledge useful discussions with W. Erwin, D. de Ceglia, and M. A. Vincenti. LRS, JT and CC acknowledge financial support from RDECOM Grant W911NF-16-1-0563 from the International Technology Center-Atlantic. The authors thank J. Sipe and M. Liscidini for critical reading of the manuscript. We thank C. Garson for his help growing most of the ITO samples.